\documentstyle[prd,aps,preprint]{revtex}
\newfont\fiverm{cmr5}
\input prepictex
\input pictex
\input postpictex

\begin{document}

\newcommand{\TeV}{\,{\rm TeV}}
\newcommand{\GeV}{\,{\rm GeV}}
\newcommand{\MeV}{\,{\rm MeV}}
\newcommand{\keV}{\,{\rm keV}}
\newcommand{\eV}{\,{\rm eV}}
\def\ap{\approx}
\def\beqar{\begin{eqnarray}}
\def\eeqar{\end{eqnarray}}
\newcommand{\bea}{\begin{eqnarray}}
\newcommand{\eea}{\end{eqnarray}}
\def\ler{\lesssim}
\def\gtr{\gtrsim}
\def\beq{\begin{equation}}
\def\eeq{\end{equation}}
\def\haf{\frac{1}{2}}
\def\plb#1#2#3#4{#1, Phys. Lett. {\bf #2B} (#4) #3}
\def\plbb#1#2#3#4{#1 Phys. Lett. {\bf #2B} (#4) #3}
\def\npb#1#2#3#4{#1, Nucl. Phys. {\bf B#2} (#4) #3}
\def\prd#1#2#3#4{#1, Phys. Rev. {\bf D#2} (#4) #3}
\def\prl#1#2#3#4{#1, Phys. Rev. Lett. {\bf #2} (#4) #3}
\def\mpl#1#2#3#4{#1, Mod. Phys. Lett. {\bf A#2} (#4) #3}
\def\rep#1#2#3#4{#1, Phys. Rep. {\bf #2} (#4) #3}
\def\lpp{\lambda''}
\def\ccg{\cal G}
\def\slash#1{#1\!\!\!\!\!/}
\def\rpv{\slash{R_p}}

\setcounter{page}{1}
\draft
\preprint{KAIST-TH 99/03, hep-ph/9902292}

\title{String or $M$ Theory Axion as a  Quintessence}

\author{Kiwoon Choi}

\address{Department of Physics,
Korea Advanced Institute of Science and Technology\\
        Taejon 305-701, Korea}

\tighten
\maketitle

\begin{abstract}
A slow-rolling scalar field ($Q\equiv$ Quintessence) with potential
energy $V_Q\sim (3\times 10^{-3} \, {\rm eV})^4$ has been proposed
as the origin of accelerating universe at present.
We investigate the effective potential of $Q$ in the framework of supergravity
model including the quantum corrections induced by generic (nonrenormalizable) 
couplings of $Q$ to the gauge and charged matter multiplets.
It is argued that
the K\"ahler potential, superpotential
and gauge kinetic functions of the underlying supergravity
model are required to be invariant under the variation of $Q$
with an extremely fine accuracy in order to provide a 
working quintessence potential.
Applying these results  for string or $M$-theory, we point out  that the 
heterotic $M$-theory or  Type I string axion can be a  plausible candidate 
for  quintessence if  (i) it does not couple to the
instanton number of gauge interactions not  weaker than those of 
the standard model 
and (ii) the modulus partner ${\rm Re}(Z)$ of the 
periodic quintessence axion ${\rm Im}(Z)\equiv {\rm Im}(Z)+1$ has
a large VEV: 
${\rm Re}(Z)\sim \frac{1}{2\pi}\ln(m_{3/2}^2 M_{Planck}^2/V_Q)$.
It is stressed that such a  large ${\rm Re}(Z)$
gives  the gauge unification scale at around the 
phenomenologically favored value $3\times 10^{16}$ GeV.
To provide an accelerating universe,
the  quintessence axion should be 
at near the top of its effective potential at present, which requires
a severe fine tuning of the initial condition of $Q$ and $\dot{Q}$
in the early universe.
We discuss a late time inflation scenario based on 
the modular and CP invariance of the
moduli effective potential,   yielding
the required initial condition  in a natural manner if the K\"ahler metric 
of the quintessence axion superfield  receives a sizable nonperturbative contribution.
\end{abstract}



\section{introduction and summary}

Recent measurements of the luminosity-red shift relation for Type Ia supernovae
suggest that the Universe is accelerating and thus
a large fraction of the energy density  has negative
pressure \cite{accel}. A nonvanishing cosmological constant (vaccum energy density) 
would be the simplest form of dark energy density providing the necessary negative pressure.
However there is a widespread  prejudice that the correct, if  exists any, 
solution to the cosmological constant problem \cite{weinberg} will lead to 
an exactly vanishing vacuum energy density.
For instance, within the framework of the Euclidean wormhole  mechanism
\cite{coleman} which has been proposed
sometime ago as a solution to the cosmological constant problem,
the vacuum energy density appears to  exactly vanish 
if all fields (except for the spacetime metric) were settled down
at their true vacuum expectation values (VEVs). 

Recently quintessence in the form of  slow-rolling scalar field 
has been proposed as an alternative form of dark energy density with negative pressure
\cite{quint}.
The true minimum of the quintessence potential is presumed to vanish, i.e.
$(V_Q)_{\rm min}=0$, 
however the present value of $Q$ is displaced from the true minimum,
providing 
\beq
V_Q\sim (3\times 10^{-3} \, {\rm eV})^4,
\label{quintpotential}
\eeq
and a negative pressure with the equation of state:
\beq
\omega=\frac{p_Q}{\rho_Q}=\frac{\frac{1}{2}{\dot{Q}}^2-V_Q}{
\frac{1}{2}{\dot{Q}}^2+V_Q} \ler -0.5,
\eeq
while obeying its equation of motion:
\beq
\ddot{Q}+3H\dot{Q}+\frac{\partial V_Q}{\partial Q}=0.
\eeq
In order to have negative pressure, we then need 
the following slow-roll condition:
\beq
\left|\frac{\partial V_Q}{\partial Q}\right|\ler 
\left|\frac{V_Q}{M_P}\right|,
\label{slowroll}
\eeq
where $M_P= M_{Planck}/\sqrt{8\pi}
=2.4\times 10^{18}$ GeV denotes the reduced Planck scale.
This slow-roll condition indicates that
the typical range of $Q$ is of order $M_P$
or at least {\it not} far below $M_P$.

Given the assumption of $(V_Q)_{\rm min}=0$,
one can in principle compute the quintessence potential $V_Q$ 
once the particle physics model for $Q$ is given.
However the required size of $V_Q$ is extremely small compared
to any of the  mass scales of particle physics,
and thus an utmost question is how such a small $V_Q$
can arise from  realistic particle physics models.
It must be stressed that $V_Q$ is required to be small {\it over}
a range of $Q$ which is essentially of order $M_P$, {\it not only}
for a particular value of $Q$.
As a result, unless there were  a symmetry 
which would enforce $V_Q$ to be flat enough,
$V_Q$ suffers from the fine tuning problem much more severe than the
conventional cosmological constant problem associated with the 
smallness of the potential energy at its minimum \cite{lyth}.
In fact, the most natural candidate for a light scalar field whose typical
range of variation is of order $M_P$ is the string or $M$-theory moduli
multiplets describing the (approximately) degenerate 
string or $M$-theory vacua.
It is thus quite tempting  to look at the possibility that $Q$
corresponds to a certain combination of
the string or $M$-theory moduli superfields.
In this paper, we wish to explore this possibility
and point out that the heterotic $M$-theory or  
Type I string axion can be a  plausible candidate for quintessence
if its modulus component has a large VEV and its K\"ahler metric
receives a sizable nonperturbative contribution.

To make the discussion more rigorous, we provide  in  section II 
a careful analysis of the low energy effective potential of 
generic quintessence\footnote{Some aspects of the
quintessence potential was discussed recently in \cite{lyth}.}
in the framework of 4-dimensional effective supergravity model 
which is presumed to describe the dynamics of $Q$ at high energy scales.
We include both the perturbative and nonperturbative corections to $V_Q$ 
induced by generic (nonrenormalizable) couplings of $Q$ 
with the gauge and charged matter multiplets in the model.
It is then  argued that
the K\"ahler potential ($K$), superpotential
($W$) and gauge kinetic functions ($f_a$) of the underlying supergravity
model are required to be invariant under the variation of $Q$
with an extremely fine accuracy, which  implies as one of
its consequences that non-derivative couplings of $Q$ are extremely
suppressed and so there is  no $Q$-mediated long range force.  Also   estimated are 
the size of $Q$-invariance breaking
for various terms in $K$, $W$ and $f_a$
which would produce the correct value of
$V_Q$.

Applying these results  for string or $M$-theory, we show  in section III  
that the heterotic $M$-theory  or  
Type I string axion can be a  plausible candidate for quintessence.
The quintessence axion may correspond to the heterotic
$M$-theory axion arising from the three form field of 11-dimensional
supergravity on a manifold with boundary \cite{horava} or 
to the Type I string axion
arising from the R-R two form field \cite{polchinski}.
To avoid a too large potential energy,
it is required that the quintessence axion does
{\it not} couple to the instanton number
of gauge interactions
{\it not} weaker than those of the standard model, particularly not to
those of the QCD and
even stronger hidden sector gauge interactions.
Then the effective potential of heterotic $M$-theory axion
is mainly induced by the membrane 
instantons wrapping the 2-cycle of the internal 6-manifold
and stretched along the 11-th dimension,
while that of Type I string axion is induced  by   $D1$ or $D5$
instantons wrapping the 2 or 6-cycle.
The resulting axion potential is estimated as 
\beq
V_Q\sim e^{-2\pi\langle {\rm Re}(Z)\rangle} m_{3/2}^2M_P^2
\,\cos [2\pi {\rm Im}(Z)],
\label{axionpotential}
\eeq
where ${\rm Re}(Z)$ is the modulus partner of the periodic quintessence
axion ${\rm Im}(Z)\equiv {\rm Im}(Z)+1$ and $m_{3/2}$ is the
gravitino mass.
Thus the quintessence potential energy (\ref{quintpotential}) is obtained
if (i) ${\rm Re}(Z)$ takes a VEV significantly larger than the self dual
value of order unity:
\beq
\langle {\rm Re}(Z)\rangle
\sim\frac{1}{2\pi}\ln (m_{3/2}^2M_P^2/V_Q)
\sim 32, 
\label{modulusvalue}
\eeq
and (ii) (one-loop) threshold corrections to 
the gauge coupling constants not weaker than those of 
the standard model are {\it $Z$-independent}, 
which would assure that the quintessence  axion does {\it not} couple to 
the instanton numbers  generating a too large 
$V_Q$.\footnote{Of course this is true only when
the tree level gauge coupling constants
are also $Z$-independent.}

As will be discussed in section IV,
when combined with $\alpha_{GUT}=1/25$ and $M_P=2.4\times 10^{18}$ GeV,
the quintessence axion potential (\ref{axionpotential})
which is presumed to take the value of (\ref{quintpotential})
fixes all the couplings and scales of the underlying model.
Particularly it gives 
the gauge unification scale
\beq
M_{GUT}
\sim  1.3\gamma\times 10^{16} \left(\frac{32}{\langle {\rm Re}(Z)\rangle}\right)^{1/2}
\, {\rm GeV},
\eeq
where $\gamma$ is a model-dependent constant of order unity
which will be defined later.
It is then interesting to note that
$\langle {\rm Re}(Z)\rangle\sim 32$ required for  a quintessence
axion determines  $M_{GUT}$ at a value 
close to the phenomenologically
favored  $3\times 10^{16}$ GeV.

One potential problem  of the string or $M$-theoretic  quintessence 
axion is that it does {\it not}   satisfy the second slow-roll
condition:
\beq
\left|\frac{\partial^2 V_Q}{\partial Q^2}\right|\lesssim 
\left|\frac{V_Q}{M_P^2}\right|.
\label{secondslow}
\eeq
As is well known, for the case of inflation in the early universe,
this additional slow-roll condition
is essential for the accelerated expansion goes over many Hubble times.
Unless this condition  is satisfied,
the quintessence axion can lead to an accelerating universe at present
{\it only} when it has a very particular initial value in the early universe,
i.e. the quintessence axion suffers from  the  fine-tuning problem
of initial condition. 
As we will see, the canonical quintessence axion field is given by
\beq
Q=M_P\sqrt{2\langle K_0^{\prime\prime}\rangle} {\rm Im}(Z),
\eeq
where 
$$
K_0^{\prime\prime}= \frac{\partial^2K_0}{\partial Z\partial Z^*}
$$
denotes the K\"ahler metric of the
quintessence axion superfield $Z$.
The  slow-roll condition (\ref{slowroll}) 
applied for the axion potential 
(\ref{axionpotential})
then leads to
\beq
|2\pi {\rm Im}(Z)|_{\rm present}\ler 
\frac{1}{2\pi} \sqrt{\langle 2K_0^{\prime\prime}\rangle}\ll 1
\label{present}
\eeq
since it is expected that $\langle K_0^{\prime\prime}\rangle \ll 1$ for
the modulus VEV (\ref{modulusvalue}). 
This implies that at present
the angular field $2\pi {\rm Im}(Z)$ takes a value at near
the top of its effective potential. 
Since $V_Q$ violates the second slow-roll
condition (\ref{secondslow}), viz 
\beq
 \left|\frac{\partial^2 V_Q}{\partial Q^2}\right|=
\left|\frac{2\pi^2}{\langle K_0^{\prime\prime}\rangle}\frac{V_Q}{M_P^2}
\right| \gg \left|\frac{V_Q}{M_P^2}\right|,
\eeq
and also ${\rm Im}(Z)=0$ corresponds to the maximum of $V_Q$,
one needs a severe fine tuning
of the initial values of $Q$ and $\dot{Q}$
in order to have the  present value satisfying (\ref{present}).
It turns out  that the degree of fine tuning is quite sensitive to 
$\langle K_0^{\prime\prime}\rangle$:
\beq
|2\pi {\rm Im}(Z)|_{\rm initial}\ler \exp (-0.5 m_0/H_0)\sim 
\exp (-3/\sqrt{\langle K_0^{\prime\prime}\rangle}),
\label{initialcondition}
\eeq
where $H_0\approx V_Q/3M_P^2$ denotes 
the Hubble expansion rate at {\it present}
and $m_0^2=|\partial^2V_Q/\partial Q^2|=|2\pi^2V_Q/M_P^2\langle K_0^{\prime\prime}\rangle|$.
If the K\"ahler metric of $Z$ is dominated by the perturbative contribution,
we have $\sqrt{\langle K_0^{\prime\prime}\rangle}\approx 1/\langle{\rm Re}(Z)\rangle\approx 3\times 10^{-2}$.
Such small  value of the K\"ahler metric would lead to too severe fine tuning of the initial value
of the quintessence axion, particularly when the quantum fluctuation of the axion
field in the early universe is  taken into account.
However there can be a large {\it nonperturbative} contribution to the K\"ahler potential \cite{banks1},
which may be necessary to stabilize ${\rm Re}(Z)$ at the value $\sim 32$ which is significantly
bigger than the self dual value $\sim 1$. 
If the stabilization of ${\rm Re}(Z)$ is achieved by the nonperturbative effects encoded
in the K\"ahler potential, the K\"ahler metric $K_0^{\prime\prime}$ would be dominated
by the nonperturbative contribution, so that $\sqrt{\langle K_0^{\prime\prime}\rangle}$
can be significantly  bigger than the perturbative value $3\times 10^{-2}$.
In this case, the degree of the fine tuning of 
the  initial conditions can be ameliorated
enough to allow a cosmological scenario  yielding  the required initial
condition (\ref{initialcondition}).
At the end of section IV, we  discuss a late time inflation  scenario based on
the modular and CP invariance of the moduli effective potential,
yielding
the initial value of the quintessence axion satisfying (\ref{initialcondition})
if the K\"ahler metric is enhanced by nonperturbative effects
as 
\beq
\sqrt{\langle K_0^{\prime\prime}\rangle}
\approx  9\times 10^{-2}.
\label{kahlerenhance}
\eeq
This scenario utilizes the specific property of 
string/M-theoretic quintessence axion ${\rm Im}(Z)$ that its potential
is  of the order of $\bar{V}e^{-2\pi {\rm Re}(Z)}$ where
$\bar{V}={\cal O}(H^2M_P^2)$ over the period during which
the expansion rate $H\gtrsim m_{3/2}$ and
$\bar{V}={\cal O}(m_{3/2}^2M_P^2)$ for the subsequent period with
$H\ll m_{3/2}$. 
This allows that ${\rm Im}(Z)$ has a mass of the order of $H$ 
during the late inflation with $H\sim m_{3/2}$ and ${\rm Re}(Z)\sim 1$,
while it becomes extremely light after the inflation
since ${\rm Re}(Z)$ is rolled down to a large value at the end
of inflation.
So during the late inflation, ${\rm Im}(Z)$ rolls toward the minimum
of the potential. If the axion potential is almost CP invariant,
which would be the case when the scale of CP violation is low enough 
compared to $M_P$,  the minimum of the axion potential during
the late inflation would coincide with either the minimum or the maximum
of the exponentially suppressed axion potential at the present.
If it coincides to the maximum and also the K\"ahler metric is
enhanced as (\ref{kahlerenhance}), the latetime inflation can provide 
the desired initial condition (\ref{initialcondition}) 
without any conflict with the known observational results.

\section{effective potential of generic quintessence}

Our starting point is the 4-dimensional $N=1$ effective supergravity 
model \cite{nillesreview} which is presumed to describe the dynamics of $Q$ at
high energy scales not far below $M_P$.
The  K\"ahler potential,
superpotential and gauge kinetic functions of the model  can be written as
\beqar
K&=& K_0(Z_i,Z_i^*)+Z_{\alpha\bar{\beta}}(Z_i,Z^*_i)C^{\alpha}C^{*\beta}+
[\frac{1}{2}H_{\alpha\beta}(Z_i,Z_i^*)C^{\alpha}C^{\beta}
\nonumber \\
&&+\frac{1}{2}X_{\alpha\beta\bar{\gamma}}(Z_i,Z^*_i)C^{\alpha}C^{\beta}C^{*\gamma}+
{\rm h.c.}]+...,
\nonumber \\
W&=&W_0(Z_i) +\frac{1}{3!}Y_{\alpha\beta\gamma}(Z_i)C^{\alpha}C^{\beta}C^{\gamma}
+\frac{1}{4!}\Gamma_{\alpha\beta\gamma\delta}(Z_i)C^{\alpha}C^{\beta}C^{\gamma}C^{\delta}+...,
\nonumber \\
f_a&=&f_{a0}(Z_i)+\frac{1}{2}F_{a\alpha\beta}(Z_i)C^{\alpha}C^{\beta}+...,
\label{sugramodel}
\eeqar
where $C^{\alpha}$ denote the gauge-charged light matter superfields whose VEVs 
are far below  $M_P$ 
and $Z_i$ stand for generic gauge-singlet light  multiplets including the fields  
with large VEVs of 
${\cal O}(M_P)$.
$Z_i$ may  correspond to the string or $M$-theory moduli superfields
and/or some gauge-invariant composite superfields made of strongly interacting  hidden sector 
fields and/or others whose typical range of variation is of ${\cal O}(M_P)$.
The quintessence $Q$ is included in the model as a certain combination of $Z_i$ and $Z^*_i$: 
\beq
Q\in \{Z_i,Z_i^*\}.
\eeq
The ellipses stand for the terms of higher order in $C^{\alpha}$, and
the leading non-renormalizable terms (in  power countings of $C^{\alpha}$)
are explicitly written to illustrate their effects  on 
$V_Q$.  
Note that unless specified 
we are using the standard supergravity unit with $M_P=1$ throughout this paper.

The superpotential $W_0(Z_i)$  is assumed to be induced
by some (nonperturbative) supersymmetry (SUSY) breaking dynamics,  e.g. hidden sector gaugino 
condensation, which is already integrated out.
The auxiliary $F$-components of certain $Z_i$ develop non-zero VEVs:
$$
F_i=\langle e^{K_0/2}D_iW_0\rangle={\cal O}(m_{3/2}M_P),
$$
thereby breaking SUSY, where $D_iW_0=\partial_iW_0+W_0\partial_iK_0$ 
is the K\"ahler covariant derivative
with $\partial_i=\partial/\partial Z_i$.
Here  we concentrate on the supergravity-mediated SUSY breaking scenario
\cite{nillesreview}
yielding the soft SUSY breaking parameters (the
soft scalar masses, gaugino masses, e.t.c.) of 
order the gravitino mass $m_{3/2}=\langle e^{K_0/2}|W_0|\rangle$.
For the case of  gauge-mediation \cite{gaugemediation}, the discussion will be 
modified but still leads to the same conclusion that
$K$, $W$, and $f_a$  should be invariant under the
variation of quintessence, i.e. $Q$-independent, 
with an extremely fine accuracy.
Note that unless constrained by some symmetries
the coefficients functions in $K$, $W$, and $f_a$
are generically $Q$-dependent,
particularly when the terms of arbitrary order in $Q$ are taken into account.
Those coefficient functions determine the (nonrenormalizable)
couplings of $Q$ to the light gauge and charged matter multiplets.
For instance, $Q^nF^a_{\mu\nu}F^{a\mu\nu}/M_P^n$ arises from  
a $Q$-dependent ${\rm Re}(f_a)$,
$Q^nF^a_{\mu\nu}\tilde{F}^{a\mu\nu}/M_P^n$ from a $Q$-dependent ${\rm Im}(f_a)$,
$Q^nC^{\alpha}\psi^{\beta}\psi^{\gamma}/M_P^n$ from a $Q$-dependent $Y_{\alpha\beta\gamma}$,
and so on.
At any rate, once the $Q$-invariance condition is fulfilled,
non-derivative couplings of $Q$ are highly suppressed, and as a result
$Q$ does not mediate any macroscopic force.

Obviously the tree level potential of $Q$ can be determined by $K_0$ and $W_0$
via the conventional supergravity potential:
\beq
V_0=e^{K_0}[K^{i\bar{j}}D_iW_0D_{\bar{j}}
W_0^*-3|W_0|^2]
\sim m_{3/2}^2M_P^2,
\label{sugratree}
\eeq
where $K^{i\bar{j}}$ is the inverse
of the K\"ahler metric $K_{ij}=\partial_i\partial_{\bar{j}}K_0$
and the typical range of $V_0$ under the variation of $Z_i={\cal O}(M_P)$
is estimated to be of order  $m_{3/2}^2M_P^2$.
When quantum corrections are taken into account,
generic (nonrenormalizable) couplings of $Q$ to charged gauge and matter multiplets  contribute
to $V_Q$. 
The leading quantum correction comes from the quadratically divergent
one-loop potential \cite{sugraoneloop} including the piece
depending on $Z_{\alpha\bar{\beta}}$
and ${\rm Re}(f_a)$:
\beq
\delta V_1=\frac{\Lambda^2}{16\pi^2}
e^{K_0}D_iW_0D_{\bar{j}}W_0^*R^{i\bar{j}}
\sim \frac{1}{16\pi^2}m_{3/2}^2\Lambda^2
\eeq
where the cutoff scale $\Lambda$ is of order the messenger
scale of SUSY breaking, which is essentially of ${\cal O}(M_P)$
for supergravity-mediated SUSY-breaking models,
\beq
R^{i\bar{j}}=K^{i\bar{k}}K^{l\bar{j}}
\partial_l\partial_{\bar{k}}[\ln(\det Z_{\alpha\bar{\beta}})
-\ln(\det{\rm Re}
(f_{ab}))],
\eeq
where $f_{ab}=f_a\delta_{ab}$, and the typical range of $\delta V_1$ is estimated
to be of order $\frac{1}{16\pi^2}m_{3/2}^2\Lambda^2$ for the variation of $Z_i={\cal O}(M_P)$.
This piece of  one-loop potential arises from
the well-known 
$V_1=\frac{1}{16\pi^2}\Lambda^2{\rm Str}(M^2)$ 
where the mass matrix $M^2$ includes the soft SUSY breaking scalar and gaugino masses
depending on $Z_{\alpha\bar{\beta}}$ and ${\rm Re}(f_a)$
\cite{sugraoneloop}.

$H_{\alpha\beta}$ and  $X_{\alpha\beta\bar{\gamma}}$ in
$K$ and the Yukawa couplings $Y_{\alpha\beta\gamma}$ in $W$
affect the potential energy at two-loop order.
It has been noted that the two-loop supergraph of Fig. 1 leads to
the quadratically divergent Yukawa-dependent potential energy \cite{bagger}:
\beq
\delta V_2\sim \frac{\Lambda^2}{(16\pi^2)^2}
e^{K_0}|DW_0|^2|Y|^2
\sim \frac{|Y|^2}{(16\pi^2)^2}m_{3/2}^2\Lambda^2,
\eeq
where
\beqar
&&|DW_0|^2= 
K^{i\bar{j}}D_iW_0D_{\bar{j}}W_0^*,
\nonumber \\
&&|Y|^2=Z^{\alpha\bar{\alpha}}Z^{\beta\bar{\beta}}
Z^{\gamma\bar{\gamma}}Y_{\alpha\beta\gamma}Y^*_{\bar{\alpha}\bar{\beta}\bar{\gamma}}.
\nonumber
\eeqar
for  $Z^{\alpha\bar{\beta}}$ being the inverse of 
$Z_{\alpha\bar{\beta}}$.
It has been noted also that the K\"{a}hler metric ($Z_{\alpha\bar{\beta}}$) and
soft SUSY-breaking scalar mass ($m^2_{\alpha\bar{\beta}}$) of $C^{\alpha}$ receive 
quadratically divergent corrections depending upon $H_{\alpha\beta}$
and $X_{\alpha\beta\bar{\gamma}}$ \cite{choileemunoz}:
\beqar 
\delta Z_{\alpha\bar{\beta}}&=&
-\frac{\Lambda^2}{16\pi^2}R_{\alpha\bar{\beta}},
\nonumber \\
\delta m^2_{\alpha\bar{\beta}}&=&\frac{\Lambda^2}{16\pi^2}e^{K_0}|W_0|^2R_{\alpha\bar{\beta}},
\eeqar
where 
$$
R_{\alpha\bar{\beta}}=Z^{\beta\bar{\alpha}}Z^{\gamma\bar{\gamma}}X_{\alpha\beta\bar{\gamma}}
X^*_{\bar{\alpha}\bar{\beta}\gamma}-
Z^{\gamma\bar{\gamma}}K^{i\bar{j}}\partial_{\bar{j}}H_{\alpha\gamma}
\partial_iH^*_{\bar{\beta}\bar{\gamma}}.
$$
When combined together as the supergraph of Fig. 2, 
these one-loop corrections lead to the following
quartically-divergent two-loop correction to the potential energy:
\beqar
\delta V_2^{\prime}&\sim&\left(\frac{\Lambda^2}{16\pi^2 }\right)^2
e^{K_0}|W_0|^2 
(|X|^2+|\bar{\partial}H|^2)
\nonumber \\
&\sim& \frac{|X|^2+|\bar{\partial} H|^2}{(16\pi^2)^2}
\frac{m_{3/2}^2\Lambda^4}{M_P^2},
\eeqar
where 
\beqar
|X|^2&=&Z^{\alpha\bar{\alpha}}Z^{\beta\bar{\beta}}Z^{\gamma\bar{\gamma}}
X_{\alpha\beta\bar{\gamma}}X^*_{\bar{\alpha}\bar{\beta}\gamma}
\nonumber \\
|\bar{\partial}H|^2&=&Z^{\alpha\bar{\alpha}}Z^{\beta\bar{\beta}}K^{i\bar{j}}
\partial_{\bar{j}}H_{\alpha\beta}\partial_iH^*_{\bar{\alpha}\bar{\beta}}.
\nonumber 
\eeqar

Finally $\Gamma_{\alpha\beta\gamma\delta}$ and $F_{a\alpha\beta}$
contributes to the potential energy through the 3-loop supergraph of Fig. 3:
\beqar
\delta V_3 &\sim& \frac{1}{(16\pi^2)^3}e^{K_0}
(\Lambda^4 |DW_0|^2|\Gamma|^2+\Lambda^6|\bar{D}W_0^*\partial F|^2)
\nonumber \\
&\sim&\frac{1}{(16\pi^2)^3}\left(\frac{m_{3/2}^2\Lambda^4}{M_P^2}|\Gamma|^2
+\frac{m_{3/2}^2\Lambda^6}{M_P^4}|\partial F|^2\right),
\eeqar
where $(\bar{D}W_0^*\partial F)_{\alpha\bar{\beta}}=K^{i\bar{j}}\partial_iF_{a\alpha\bar{\beta}}
D_{\bar{j}}W_0^*$ and
\beqar
&&|\Gamma|^2=
Z^{\alpha\bar{\alpha}}Z^{\beta\bar{\beta}}Z^{\gamma\bar{\gamma}}Z^{\delta
\bar{\delta}}\Gamma_{\alpha\beta\gamma\delta}\Gamma^*_{\bar{\alpha}
\bar{\beta}\bar{\gamma}\bar{\delta}},
\nonumber \\
&&|\partial F|^2=
K^{i\bar{j}}Z^{\alpha\bar{\beta}}Z^{\beta\bar{\alpha}}\partial_iF_{a\alpha\beta}
\partial_{\bar{j}}F^*_{a\bar{\alpha}\bar{\beta}}.
\eeqar

So far, we have discussed the effective potential  induced by
the coefficient functions  
in $K$, $W$ and $f_a$ except for ${\rm Im}(f_{0a})$
whose VEV corresponds to the vacuum angle of the $a$-th gauge group.
In the effective supergravity model under consideration, 
possible nonperturbative hidden sector gauge interactions
were already integrated out, and their effects are encoded in
the effective superpotential $W_0$ which is a function of 
the holomorphic hidden sector gauge kinetic functions.
This means that the effective potential 
of the both real and imaginary parts of the
gauge kinetic functions of strongly interacting  hidden sector
are already included in (\ref{sugratree}).

Let us now  consider the effective potential
of the vacuum angles of the standard model gauge group. 
Due to the asymtotic freedom,
the potential energy of the QCD vacuum angle $8\pi^2{\rm Im}(f_{QCD})$
is induced mainly by the low energy QCD dynamics at 
energy scales around 1 GeV \cite{kimreview}, and thus  is unambiguously determined as
\beq
V_{QCD}\sim f_{\pi}^2m_{\pi}^2\cos[8\pi^2{\rm Im}(f_{QCD})],
\label{qcdaxion}
\eeq
where $m_{\pi}$ and $f_{\pi}$ denote the pion mass and decay constant, respectively.
However the potential energy  of the  electroweak vacuum angle
$8\pi^2{\rm Im}(f_{EW})$ is mainly due to the $SU(2)_L$ instantons
at high energy scales around $M_{GUT}$.
Since the multi-fermion vertex of the standard model fermions  induced 
by the electroweak instanton violates both the baryon ($B$) and lepton
($L$) numbers,
$\triangle B=\triangle L=3$,
the resulting  potential energy is suppressed by
the insertions of small $B$ and $L$-violating couplings
in addition to the suppressions  by  the semiclassical factor $e^{-2\pi/\alpha_{GUT}}$
and the small SUSY breaking factor of order $m_{3/2}^2$.
We thus have
\beq
V_{EW}\sim \epsilon e^{-2\pi/\alpha_{GUT}}m_{3/2}^2M_P^2
\cos[{\rm Im}(8\pi^2f_{EW})],
\label{ewaxion}
\eeq
where the small factor $\epsilon$ includes $B$ and $L$ violating couplings
obeying the selection rule $\triangle B=\triangle L=3$ together with
small Yukawa-type couplings and loop factors which are necessary to close all
fermion zero modes \cite{hyungdo}.
The size of $\epsilon$ is highly model-dependent, particularly
on the $B$, $L$, and fermion chirality-changing couplings available 
at $M_{GUT}$.

Summing  all the contributions discussed so far, the typical size of
the quintessence potential $V_Q$ 
is estimated to be
\beqar
V_Q 
&\sim&
m_{3/2}^2M_P^2\left[\frac{\delta \ln (K_0)}{
\delta Q}+\frac{\delta \ln (W_0)}{\delta Q}
+\frac{\Lambda^2}{16\pi^2 M_P^2}\left(
\frac{\delta Z_{\alpha\bar{\beta}}}{\delta Q}+\frac{\delta {\rm Re}(f_a)}{\delta Q}\right)
\right.\nonumber \\
&&
+\frac{\Lambda^2}{(16\pi^2)^2M_P^2}\frac{\delta |Y|^2}{\delta Q}
+\frac{\Lambda^4}{(16 \pi^2)^2M_P^4}
\left(\frac{\delta |X|^2}{\delta Q}+\frac{\delta |\bar{\partial}H|^2}{\delta Q}
\right)
+\frac{\Lambda^4}{(16\pi^2)^3M_P^4}\frac{\delta |\Gamma|^2}{\delta Q}
\nonumber \\
&& 
\left.
+\frac{\Lambda^6}{(16\pi^2)^3M_P^6}\frac{\delta |\partial F|^2}{\delta Q}
+\frac{f_{\pi}^2m_{\pi}^2}{m_{3/2}^2M_P^2}\frac{\delta {\rm Im}(f_{QCD})}{\delta Q}
+\epsilon e^{-2\pi/\alpha_{GUT}}\frac{\delta {\rm Im}(f_{EW})}{\delta Q}
\right]Q_{\rm typ},
\label{sugraquintpotential}
\eeqar
where 
$\delta G$ denotes the variation of the coefficient function
$G$ for the quintessence variation $\delta Q$
and $Q_{\rm typ}$ is the typical range of the quintessence
variation. Note that (\ref{sugraquintpotential}) represents 
the variation of $V_Q$ under the variation of $Q$
specified by $Q_{\rm typ}$. 
In order that $V_Q\sim (3\times 10^{-3} \, {\rm eV})^4$
over the range of $Q_{\rm typ}$,
its variation should not  significantly exceed 
$(3\times 10^{-3} \, {\rm eV})^4$
over the same  range.
This requires that each term in (\ref{sugraquintpotential}) 
does not   significantly exceed $(3\times 10^{-3} \, {\rm eV})^4$
over the range of $Q_{\rm typ}$.
Note that although there may be a significant cancellation
between different terms for a particular value of $Q$,
it is extremely difficult that the cancellation persists
over the range of $Q_{\rm typ}$ 
which is essentially of order $M_P$ \cite{lyth}.
Requiring each term in (\ref{sugraquintpotential}) not exceed  
$(3\times 10^{-3} \, {\rm eV})^4$,
we find
\beqar
&&\frac{\delta \ln (K_0)}{\delta Q}\ler \frac{10^{-88}}{\kappa_Q\kappa_{3/2}},
\quad 
\frac{\delta \ln (W_0)}{\delta Q}\ler \frac{10^{-88}}{\kappa_Q\kappa_{3/2}},
\nonumber \\
&&\frac{\delta  Z_{\alpha\bar{\beta}}}{\delta Q}\ler 
\frac{10^{-81}}{\kappa_Q\kappa_{\Lambda}\kappa_{3/2}},
\quad
\frac{\delta |\bar{\partial}H|^2}{\delta Q}\ler 
\frac{10^{-74}}{\kappa_Q\kappa_{\Lambda}^2\kappa_{3/2}},
\nonumber \\
&&\frac{\delta |X|^2}{\delta Q} \ler \frac{10^{-74}}{\kappa_Q\kappa_{\Lambda}^2\kappa_{3/2}},
\quad
\frac{\delta |Y|^2}{\delta Q} \ler \frac{10^{-79}}{\kappa_Q\kappa_{\Lambda}\kappa_{3/2}},
\nonumber \\
&&\frac{\delta |\Gamma|^2}{\delta Q}\ler 
\frac{10^{-72}}{\kappa_Q\kappa_{\Lambda}^2\kappa_{3/2}},
\quad
\frac{\delta |\partial F|^2}{\delta Q}\ler 
\frac{10^{-67}}{\kappa_Q\kappa_{\Lambda}^3\kappa_{3/2}},
\nonumber \\
&&\frac{\delta {\rm Re}(f_a)}{\delta Q}\ler \frac{10^{-81}}{\kappa_Q\kappa_{\Lambda}\kappa_{3/2}},
\quad
\frac{\delta {\rm Im}(f_{QCD})}{\delta Q} \ler\frac{10^{-42}}{\kappa_Q\kappa_{3/2}},
\label{constraint}
\eeqar
where $\kappa_Q=Q_{\rm typ}/M_P$, 
$\kappa_{\Lambda}=(\Lambda/10^{16} \, {\rm GeV})^2$,
$\kappa_{3/2}=(m_{3/2}^2/ {\rm TeV})^2$, 
and all other quantities are defined in the supergravity unit
with $M_P=1$, e.g. $Q$, $K_0$,  $W_0$,  and $X_{\alpha\beta\bar{\gamma}}$
correspond to $Q/M_P$, $K_0/M_P^2$,
$W_0/M_P^3$, and $M_PX_{\alpha\beta\bar{\gamma}}$,
respectively.
Note that at least one of the variations is required to 
saturate its upper limit to produce the
desired quintessence potential energy.

When expanded in powers of $C^{\alpha}$ and also 
written in the unit with $M_P=1$,
all gauge-invariant coefficient functions in $K$, $W$, and $f_a$ are 
expected to be of order unity 
except for $W_0$ which is of order $m_{3/2}$.
Furthermore none of  $\kappa_Q$, $\kappa_{\Lambda}$, and $\kappa_{3/2}$
can be small enough to significantly compensate the extremely small
numerators in (\ref{constraint}).
Then the limits of (\ref{constraint})  
on the variations of coefficient functions
imply that $K$, $W$ and $f_a$ {\it must be invariant under the variation 
of $Q$, i.e. $Q$-independent, with an extremely fine accuracy.}
Note that terms of  higher order in $C^{\alpha}$
which are not explicitly discussed here
are similarly constrained
to be $Q$-independent.
At any rate, if this $Q$-invariance is satisfied,
$Q$ behaves like a (pseudo) Goldstone boson having only {\it derivative} couplings
and thus does {\it not} mediate any macroscopic force.

A possible breaking of the $Q$-invariance whose size was
not estimated in (\ref{constraint}) 
is the variation of the electroweak vacuum angle $8\pi^2{\rm Im}(f_{EW})$ 
which is constrained
as
\beq
\frac{\delta {\rm Im}(f_{EW})}{\delta Q} \ler\frac{10^{-20}}{\epsilon\kappa_Q\kappa_{3/2}}.
\eeq
Although quite sensitive to the physics at $M_{GUT}$, generically
$\epsilon$ is suppressed by many powers of small Yukawa couplings,
loop factors, and also small $B$ and/or $L$-violating couplings,
and so it can be easily smaller than $10^{-20}$ \cite{hyungdo}.
This would suggest that ${\rm Im}(f_{EW})$ can have a sizable
$Q$-dependence,  and so a sizable $Q$-coupling to the electromagnetic 
$F_{\mu\nu}\tilde{F}^{\mu\nu}$ which may lead to an interesting
observable consequence \cite{carroll}.
However in view of the gauge coupling unification,
i.e. $\langle{\rm Re}(f_{QCD})\rangle=
\langle{\rm Re}(f_{EW})\rangle$, it is highly unlikely that
${\rm Im}(f_{EW})$ has a sizable $Q$-dependence, while ${\rm Im}(f_{QCD})$
is $Q$-independent as in (\ref{constraint}).\footnote{A possible 
loophole \cite{kim} for this argument is the possibility
of massless up quark which has been argued
to be phenomenologically viable \cite{upquark}.
In this case, the effective potential of the QCD vacuum angle
${\rm Im}(f_{QCD})$ is suppressed by $z=m_u/m_d$, i.e.
$V_{QCD}\sim zf_{\pi}^2m_{\pi}^2$, and then
the upper bound on $\delta {\rm Im}(f_{QCD})/\delta Q$ in Eq.(\ref{constraint})
should be read as $10^{-42}/z\kappa_Q\kappa_{\Lambda}$.
Thus for $|z|\ler 10^{-40}$,
${\rm Im}(f_{QCD})$ can have a sizable $Q$-dependence
and so $Q$ can couple to both $(F\tilde{F})_{QCD}$ and $(F\tilde{F})_{EW}$.}

Let us close this section with a brief discussion of
how plausible are some of the frequently used  forms of  $V_Q$
\cite{quint}
in view of our discussions above.

(a) Exponential potential: $V_Q\sim M^4 e^{-Q/f_Q}$.
This form of $V_Q$ may arise from some nonperturbative
dynamics when $Q$ corresponds to a dilaton \cite{binetruy}.
At certain level, one may  adjust the dynamics of the model 
to make this exponential potential to be of order $(3\times 10^{-3} \, {\rm eV})^4$.
However usually the gauge coupling constants in the model
are determined by the dilaton VEV, which means
$\delta {\rm Re}(f_a)/\delta Q$ is of order unity.
Then the supergravity
loop effects depending on 
$\delta {\rm Re}(f_a)/\delta Q$
(see Eq.(\ref{sugraquintpotential}))  yield a dilaton potential energy
of order $\frac{1}{16\pi^2}m_{3/2}^2\Lambda^2$ which is too large
to be a quintessence potential energy. 

(b) Inverse power law potential: $V_Q\sim M^{4+k}/Q^k$.
This form of $V_Q$ may arise from a nonperturbative dynamics when
$Q$ corresponds to a composite degree of freedom \cite{binetruy}.
An attractive feature of this potential is that $M$ does not have
to be too small compared to the particle physics mass 
scales, e.g. $M\gtr 10$ GeV 
for $k\geq 3$ and $Q\sim M_P$.
However the inverse power law behavior is so easily upset 
either by the supersymmetry breaking dynamics
yielding the soft mass term $\sim m_{3/2}^2 Q^2$
or by the Planck scale dynamics  which would generate 
a non-renormalizable term 
$\delta V_Q\sim m_{3/2}^2M_P^2Q^{n}/M_P^n$.
If any of such terms is induced, which is extremely difficult
to avoid, the variation of $V_Q$ under $\delta Q\sim M_P$
would be  too large to provide a quintessence potential.

(3) Pseudo-Goldston boson potential: $V_Q\sim M^4 \cos (Q/f_Q)$.
This form of $V_Q$ arises when $Q$ corresponds to a (pseudo) Goldstone
boson, which is perhaps the most plausible possibility
in view of the $Q$-invariance conditions of (\ref{constraint}) \cite{kim}.
Obviously then the $Q$-invariance corresponds to the nonlinearly realized
global symmetry associated with the Goldstone boson $Q$.
Still the remained  question is  how the explicit
breaking of the non-linear global symmetry can be so tiny as in (\ref{constraint}),
which is very nontivial to achieve in view of that  global symmetries
are generically broken by  Planck scale physics \cite{holman}.
In the next section, we argue that heterotic $M$-theory or Type I string axion
can be a plausible candidate for the quintessence Goldstone boson
with the tiny $Q$-invariance breaking
induced by the membrane or $D$-brane instantons.

\section{heterotic $M$
or Type I string axion  as quintessence}

The most natural candidate for a light scalar field whose typical
range of variation is of order $M_P$ is the string or $M$-theory moduli multiplets
describing the (approximately) degenerate string or $M$-theory vacua.
It is thus quite tempting  to look at the possibility that $Q$
corresponds to a certain combination of
the string or $M$-theory moduli superfields.
In this section, we examine this possibility and point out that
the heterotic $M$-theory or Type I string axion can be a plausible
candidate for quintessence.
The moduli superfields of our interests are $Z_I=(S,T_i)$
including the axion components
${\rm Im}(Z_I)$ together with
the modulus components ${\rm Re}(Z_I)$ which  correspond to
the string dilaton or the length of the 11-th segment and 
the K\"ahler moduli describing the size and shape
of the internal 6-manifold.
Explicit computations show that
generically 
\beq
\frac{\delta K_0}{\delta {\rm Re}(Z_I)}={\cal O}(1),
\eeq
and so the effective potential of the modulus components are
of order $m_{3/2}^2M_P^2$ and the modulus masses are of order
$m_{3/2}$ in view of the analysis in the previous section.
It thus appears that the modulus components
${\rm Re}(Z_I)$ can {\it not} provide a quintessence.
However as we will see still  quintessence can arise as a 
linear combination of the axion components ${\rm Im}(Z_I)$
if the corresponding  modulus component has a large VEV $\sim 32$.


Let us first consider the  heterotic $M$-theory  on a
11-dimensional manifold with boundary which is invariant
under the $Z_2$-parity \cite{horava}:
\beq
C\rightarrow -C,
\quad
x^{11}\rightarrow -x^{11},
\eeq
where $C=C_{ABC}dx^Adx^Bdx^C$ is the 3-form field in the 11-dimensional supergravity.
When compactified to 4-dimensions, axions arise
as the massless modes of $C_{\mu\nu 11}$ and $C_{mn 11}$:
\beq
\epsilon^{\mu\nu\rho\sigma}\partial_{[\nu}C_{\rho\sigma]11}
=\partial^{\mu}\eta_S,
\quad
C_{mn 11}=\sum_i \eta_i(x^{\mu}) \omega^i_{mn},
\eeq
where $\omega^i$ ($i=1$ to $h_{1,1}$) form the basis of the integer $(1,1)$ cohomology
of the internal 6-manifold and  $\mu$, $\nu$ are tangent to 
the noncompact 4-dimensional spacetime.
In 4-dimensional effective supergravity, these axions appear as 
the pseudo-scalar components of chiral multiplets:
\beqar
&&S=(4\pi)^{-2/3}\kappa^{-4/3}V+i\eta_S,
\nonumber \\
&&T_i=(4\pi)^{-1/3}\kappa^{-2/3}\int_{{\cal C}_i}\omega\wedge dx^{11}
+i\eta_i,
\label{st}
\eeqar
where $\kappa^2$ denotes the 11-dimensional gravitational coupling,
$V$ is the volume of the internal 6-manifold with the K\"ahler
two form $\omega$, and
the integral  is over the 11-th segment and   
also over the 2-cycle ${\cal C}_i$ dual to $\omega^i$.
Here the axion components are normalized by the
discrete Peccei-Quinn (PQ) symmetries:
\beq
{\rm Im}(S)\rightarrow {\rm Im}(S)+1,
\quad
{\rm Im}(T_i)\rightarrow {\rm Im}(T_i)+1,
\label{discretepq}
\eeq
which are the parts of discrete modular symmetries
involving also the dualities between large and small ${\rm Re}(S)$
or ${\rm Re}(T_i)$.

Holomorphy and the discrete PQ symmetries imply that
in the large ${\rm Re}(S)$ and ${\rm Re}(T_i)$ limits
the gauge kinetic functions can be written as\footnote{
This feature of the gauge kinetic function was first noted
in \cite{nilles}.}
\beq
4\pi f_a=k_a S+\sum_il_{ai}T_i+
{\cal O}(e^{-2\pi S}, e^{-2\pi T_i}),
\label{gaugekinetic}
\eeq
where $k_a$ and $l_{ai}$ are model-dependent {\it quantized} real constants
and the exponentially suppressed terms  are possibly
due to  the membrane or 5-brane instantons.
For a wide class of compactifed heterotic $M$-theory, 
we have \cite{banks,nilles1,maxion,benakli,stieberger}
\beqar
4\pi f_{E_8}&=& S+\sum_i l_iT_i+{\cal O}(e^{-2\pi S}, e^{-2\pi T_i}),
\nonumber \\
4\pi f_{E_8^{\prime}}&=& S-\sum_i l_i T_i
+{\cal O}(e^{-2\pi S}, e^{-2\pi T_i}),
\label{mgaugekinetic}
\eeqar
where $l_iT_i$ corresponds to the one-loop threshold
correction 
in perturbative heterotic string terminoloy
with the quantized coefficients $l_i$ determined by the instanton numbers
on the hidden wall and also the orbifold twists.\footnote{
For compactifications on  smooth Calabi-Yau space, $l_i$
can be determined either in the perturbative heterotic string limit
\cite{choikim,banks}
or in the 11-dimensional supergravity limit \cite{witten,benakli}
and they take half-integer values \cite{maxion}, 
while they can be generic rational
numbers for orbifolds \cite{stieberger}.}

Let $Q$ denote a linear combination of ${\rm Im}(T_i)$
for which the combination $\sum_i l_{i}T_i$ is {\it invariant}
under
\beq
U(1)_Q: \quad Q\rightarrow Q+{\rm constant}.
\label{Q-pqsymmetry}
\eeq
Note that such $Q$  exists always  as long as $h_{1,1}>1$.
In this regard, models of particular interest are the
recently discovered threshold-free models with $l_i=0$
\cite{stieberger} for which
any of ${\rm Im}(T_i)$ can be identified as $Q$.
At any rate, once such $Q$ exists, we have
\beq
\frac{\delta f_a}{\delta Q}={\cal O}(e^{-2\pi T}),
\label{smallgauge}
\eeq
where the internal 6-manifold is assumed to be isotropic
and thus ${\rm Re}(T_i)\ap {\rm Re}(T)$ for all $T_i$.
Holomorphy and discrete PQ symmetries
imply also
\beq
\frac{\delta Y_{\alpha\beta\gamma}}{\delta Q}={\cal O}(e^{-2\pi T})
\label{smallyukawa}
\eeq
for the Yukawa couplings (and also higher order holomorphic couplings)
in the superpotential of (\ref{sugramodel}).
If $W_0$ is induced by hidden sector dynamics described
by {\it $Q$-independent} gauge,  Yukawa, and higher order
holomorphic couplings,
we have
\beq
\frac{\delta W_0}{\delta Q}=
\frac{\delta W_0}{\delta f_a}
\frac{\delta f_a}{\delta Q}+
\frac{\delta W_0}{\delta Y_{\alpha\beta\gamma}}
\frac{\delta Y_{\alpha\beta\gamma}}{\delta Q}+...
={\cal O}(e^{-2\pi T}W_0).
\label{smallsuper}
\eeq

About the K\"{a}hler potential,
the discrete PQ symmetries (\ref{discretepq}) imply that
it can be written as
$K=\sum_n K_n \exp[i2\pi n{\rm Im}(T_i)]$ where
$K_n$ is a function of $Re(T_i)$ and other moduli (and also
of charged matter superfields). 
Obviously then $U(1)_Q$ is broken
by $K_n$ with $n\neq 0$.
To estimate the size of $K_n$,
it is convenient to consider the 4-dimensional $N=1$ vacuum as 
a deformation of $N=2$ vacuum {\it keeping} the discrete PQ symmetries
unbroken and also ${\rm Re}(T_i)$ at large values \cite{maxion}.
Note that this deformation is not a shift
within the $N=2$ moduli space, but
involves  the shifts of massive fields 
breaking $N=2$ supersymmetry down to $N=1$ supersymmetry.
At $N=2$ vacuum, $T_i$ correspond to the coordinates of a special
K\"{a}hler manifold whose K\"ahler potential is determined
by the {\it holomorphic} prepotential \cite{cremmer,curio},
and thus the discrete PQ symmetries
imply that 
$K_n\propto e^{-2\pi nT_i}$. 
As long as the discrete PQ symmetries 
are kept unbroken and also ${\rm Re}(T_i)$ kept large, 
this exponential suppression
of the $U(1)_Q$-breaking is expected to be maintained 
when $N=2$ vacuum is deformed to 
$N=1$ vacua of phenomenological interests,
implying that
\beq
\frac{\delta K}{\delta Q}={\cal O}(e^{-2\pi T})
\label{smallkahler}
\eeq
even at $N=1$ vacua.\footnote{
This conclusion is supported by
the observation that the nonlinear PQ transformation 
$U(1)_i: \delta T_i=
ic_i$ ($c_i=$ real constant) originates from
the {\it local} transformation of the 11-dimensional three form field, 
$\delta C=
\omega^i\wedge dx^{11}\partial_{11}\beta^i(x^{11})$
with $c_i=\int \delta C=\int \omega^i\wedge dx^{11} \partial_{11}\beta^i$.
As was noted in \cite{banks}, this suggests that 
$U(1)_i$  is broken {\it only} by nonperturbative effects
defined on the boundaries or by those involving
the bulk degree of freedoms wrapping the 2-cycle ${\cal C}_i$
dual to $\omega^i$
and stretched between the boundaries.
Since $U(1)_Q$ was chosen to avoid the breaking by  gauge
instantons on the boundaries,
it appears to be  broken only by 
the stretched membrane instantons wrapping ${\cal C}_i$
whose effects are suppressed by $e^{-2\pi T_i}$.}

When applied for the effective
potential analysis of the previous section, 
Eqs. (\ref{smallgauge}), (\ref{smallyukawa}),
(\ref{smallsuper}), and (\ref{smallkahler})  
assure that in the large ${\rm Re}(T)$ limit
the quintessence axion potential is exponentially suppressed as
\beq
V_Q\sim e^{-2\pi\langle {\rm Re}(T)\rangle}m_{3/2}^2M_P^2\cos[2\pi {\rm Im}(T)]
\label{maxionpotential}
\eeq
even when all possible quantum corrections are taken into account.
Thus what is necessary and sufficient for this axion potential to be of order
$(3\times 10^{-3} \, {\rm eV})^4$ is
\beq
\langle {\rm Re}(T)\rangle
\sim\frac{1}{2\pi}\ln(m_{3/2}^2M_P^2/V_Q)
\sim 32,
\label{modulusvev}
\eeq
where $m_{3/2}\sim 1$ TeV are used for numerical estimate.\footnote{
About stabilizing  ${\rm Re}(T)$ 
at this large value,  it has been shown in \cite{hdkim} that
both ${\rm Re}(S)$ and ${\rm Re}(T)$ 
can be stabilized at large VEVs of order $\frac{1}{\alpha_{GUT}}$
by the combined effects of the multi-gaugino condensations and
the membrane instantons wrapping the 3-cycle of the internal 6-manifold.}
As we will see in the next section, 
this value of ${\rm Re}(T)$ can be compatible with
$\alpha_{GUT}\sim 1/25$ {\it only} in the heterotic $M$-theory
limit, {\it not} in the perturbative heterotic string limit.
Although the above quintessence axion potential
has been obtained by the macroscopic analysis based on
supersymmetry and discrete PQ symmetries, one can easily
identify its microscopic origin by noting that
$2\pi {\rm Re}(T_i)=(4\pi)^{-1/3}\kappa^{-2/3}\int_{{\cal C}_i}\omega\wedge
dx^{11}$ corresponds to the Euclidean action of
the membrane instanton wrapping ${\cal C}_i$ and stretched along
the 11-th segment \cite{banks}.
When extrapolated to the perturbative heterotic string limit,
such membrane instanton corresponds to the heterotic string
worldsheet instanton wrapping the same 2-cycle \cite{worldsheetinstanton}.
Explicit computations then show that 
the K\"ahler potential and/or the gauge kinetic functions
are indeed corrected by worldsheet instantons, yielding 
$\delta K_0={\cal O}(e^{-2\pi T})$ and $\delta f_a={\cal O}(e^{-2\pi T})$
\cite{worldsheetinstanton1}.
These corrections  can be smoothly extrapolated back to
the heterotic $M$-theory limit and identified as the corrections induced
by stretched membrane instantons.
When integrated out, the correction to the hidden sector gauge kinetic function
leaves its trace in the effective superpotential as $\delta W_0=
{\cal O}(e^{-2\pi T}W_0)$.
The axion potential
(\ref{maxionpotential})
is then obtained from the supergravity potential 
(\ref{sugratree}) with $\delta K_0={\cal O}(e^{-2\pi T})$ and/or $\delta W_0=
{\cal O}(e^{-2\pi T}W_0)$.



Let us now turn to  Type I string axions.
The Type I axions (again normalized by
the discrete PQ symmetries of (\ref{discretepq}))
correspond to the massless
modes of the R-R two form fields $B_{\mu\nu}$ and $B_{mn}$,
$\epsilon^{\mu\nu\rho\sigma}\partial_{\nu}B_{\rho\sigma}
=\partial^{\mu}\eta_S$, $B_{mn}=\sum_i\eta_i(x^{\mu})\omega^i_{mn}$,
and form 4-dimensional chiral multiplets together with the string dilaton
$e^D$ and the internal space volume $V$:
\beqar
&&S=(2\pi)^{-6}\alpha^{\prime -3}e^{-D}V+i\eta_S,
\nonumber \\
&&T_i=(2\pi)^{-2}\alpha^{\prime -1}e^{-D}\int_{{\cal C}_i}\omega
+i\eta_i.
\eeqar
Here we include $D9$ and $D5$ branes in the vaccum configuration,
and consider the 4-dimensional gauge couplings
$\alpha_9$ and $\alpha_{5i}$ defined on
$D9$ branes wrapping the internal 6-manifold and 
$D5$ branes wrapping the 2-cycles ${\cal C}_i$, respectively
\cite{munoz}.
Again holomorphy and discrete PQ symmetries imply that the corresponding
gauge kinetic functions can be written as
(\ref{gaugekinetic}).
A simple leading order calculation gives $\alpha_9=1/{\rm Re}(S)$
and $\alpha_{5i}=1/{\rm Re}(T_i)$, and so \cite{munoz}
\beq
4\pi f_9= S, \quad
4\pi f_{5i}= T_i.
\label{leading}
\eeq
However this leading order result
can receive  perturbative and/or non-perturbative corrections.
Generic perturbative corrections can be expanded
in powers of the string coupling $e^{D}$, the
string inverse tension $\alpha^{\prime}$,
and also the inverse tension $e^D\alpha^{\prime k}$ of $D_{2k-1}$ branes
\cite{polchinski}.
Generically they scale as
\beq
e^{nD}\alpha^{\prime m}
\propto 
[{\rm Re}(T)]^{\frac{n+m}{2}}
[{\rm Re}(S)]^{-\frac{3n+m}{2}}.
\eeq
Combined with the leading order result (\ref{leading}) and also
the general form of gauge kinetic function
(\ref{gaugekinetic}) dictated by holomorphy and discrete PQ symmetries,
this scaling behavior implies that
$f_9$ can receive a $T_i$-dependent correction
at order $\alpha^{\prime 2}$, while there is no perturbative correction to $f_{5i}$,
and so
\beqar
4\pi f_9&=& S+l_{i}T_i+{\cal O}(e^{-2\pi S},e^{-2\pi T_i}),
\nonumber \\
4\pi f_{5i}&=& T_i+{\cal O}(e^{-2\pi S}, e^{-2\pi T_i}).
\label{type1gauge}
\eeqar

Similarly to the case of heterotic $M$-theory axion,
the quintessence  $Q$ can arise   as a linear combination of 
${\rm Im}(S)$ and ${\rm Im}(T_i)$, however its explicit form
depends on how the gauge couplings {\it not}
weaker than those of the standard model are embedded in the model.
Note that for such {\it not-so-weak} gauge couplings
both ${\rm Re}(f_a)$ and ${\rm Im}(f_a)$ 
are required to be $Q$-independent as in (\ref{constraint}), while for
the weaker gauge interactions with $\alpha_a(M_{GUT})\ler 1/32$,
${\rm Im}(f_a)$ are allowed to have a sizable $Q$-dependence.
Here are some  possibilities.
If all of the  not-so-weak gauge couplings are embedded in $\alpha_{5i}$,
${\rm Im}(S)$ can be a quintessence when 
$\langle {\rm Re}(S)\rangle\sim 32$ which is necessary
for $V_Q\sim (3\times 10^{-3} \, eV)^4$,
and also the associated gauge coupling
$\alpha_9(M_{GUT})\ler 1/32$ which may be necessary to avoid a too large
$V_Q$ induced by the gauge instantons of $\alpha_9$.
If  not-so-weak couplings are embedded in $\alpha_9$,
any of ${\rm Im}(T_i)$ can be a quintessence when $\langle {\rm Re}(T_i)\rangle\sim 32$. 
If the vacuum does not include any 
$D5$ brane wrapping the particular 2-cycle, e.g. the $i$-th cycle,
the corresponding axion ${\rm Im}(T_i)$ can be a quintessence
independently of the embedding when $\langle{\rm Re}(T_i)\rangle\sim 32$.

Once the quintessence component satisfying the constraints
on gauge kinetic functions is identified,
the $Q$-dependence of the K\"ahler potential and superpotential
is suppressed by $e^{-2\pi Z}$ ($Z=S$ or $T$) as in the case of heterotic $M$-theory,
and so the quintessence axion potential is again given by
(\ref{maxionpotential}) where now $T$ can be either
$S$ or $T_i$.
Microscopic origin of this Type I axion potential can be easily identified also
by noting  that $2\pi {\rm Re}(S)$ corresponds to 
the Euclidean action of $D5$ brane instanton wrapping the internal
6-manifold and $2\pi {\rm Re}(T_i)$ is of $D1$ string instanton wrapping
the 2-cycle ${\cal C}_i$.

\section{Couplings and scales with quintessence axion}

In the previous section, it was noted that 
the heterotic $M$ or Type I string axion can provide a 
quintessence dark energy when its modulus partner has a VEV
$\sim 32$. In this section, we discuss the couplings
and scales associated with  this modulus VEV.

\medskip

A. {\it Couplings and Scales in Heterotic $M$-theory}:

\medskip

For simplicity, let us consider
a model without $T$-dependent threshold corrections, i.e.
the quantized coefficients $l_i=0$ in Eq. (\ref{mgaugekinetic}).
We then have \cite{witten,banks}
\beqar
&&\frac{1}{\alpha_{GUT}}=
(4\pi)^{-2/3}\kappa^{-4/3}V={\rm Re}(S),
\nonumber \\
&&\omega=\frac{(4\pi)^{1/3}\kappa^{2/3}}{\pi\rho}\sum_i {\rm Re}(T_i)\omega^i,
\nonumber \\
&&V=\frac{1}{3!}\int \omega\wedge \omega\wedge\omega\ap
\frac{4\pi\kappa^2}{(\pi\rho)^3}\frac{\sum_{ijk} C_{ijk}}{6}[{\rm Re}(T)]^3
\eeqar
where $\pi\rho$ is the length of the 11-th segment,
$\omega$ is the K\"ahler form of the internal 6-manifold,
and $C_{ijk}=\int\omega^i\wedge\omega^j\wedge\omega^k$
are the intersection numbers of the integer $(1,1)$ cohomoloy basis $\omega^i$.
Here we have used the isotropy condition 
${\rm Re}(T_i)\ap {\rm Re}(T)$.
Combining these with 
$$
M_P^2=\pi\rho V/\kappa^2,
$$
one easily finds
\beqar
&&M_{KK}\equiv \frac{\gamma}{V^{1/6}}
=\frac{\gamma M_P}{
2\sqrt{\pi}(\sum C_{ijk}/6)^{1/6}[{\rm Re}(S)]^{1/2}[{\rm Re}(T)]^{1/2}}
\nonumber \\
&&\kappa^{-2/3}(\pi\rho)^3=(4\pi)^{1/3}\left(\frac{\sum C_{ijk}}{6}\right)
\frac{[{\rm Re}(T)]^3}{{\rm Re}(S)}
\label{mscale}
\eeqar
where $M_{KK}$ denotes the Kaluza-Klein threshold mass scale for
the internal 6-manifold with volume $V$.
Here $\gamma$ is a constant of order unity whose  precise value
depends on the details of compactification and also on
the regularization of the threshold effects due to massive Kaluza-Klein modes.
(For instance, $\gamma$ is expected to be around $2\pi$
for  toroidal compactifications, however it can be smaller for more
complicate compactifications, e.g. about $\sqrt{2\pi}$ for the compactification
on $S^6$.)

In heterotic $M$-theory, $M_{KK}$  can be
identified as the unification scale of gauge couplings on
the boundary \cite{witten}: 
$$M_{GUT}=M_{KK}.
$$
Also the VEVs of ${\rm Re}(S)$ and ${\rm Re}(T)$
can be determined by the two phenomenological inputs:
$\alpha_{GUT}=1/25$ and $V_Q= (3\times 10^{-3} \, {\rm eV})^4$,
yielding 
\beqar
&&{\rm Re}(S)=\frac{1}{\alpha_{GUT}}\sim 25,
\nonumber \\
&&{\rm Re}(T)= \frac{1}{2\pi}\ln(m_{3/2}^2M_P^2/V_Q)\sim 32.
\nonumber
\eeqar
Applying these moduli VEVs to (\ref{mscale}), 
we find
\beq
M_{GUT}=\frac{1.3\gamma}{ (\sum C_{ijk}/6)^{1/6}}\times 10^{16}
\, \, {\rm GeV},
\nonumber
\eeq
which is very close to the phenomenologically favored
value $3\times 10^{16}$ GeV.  We stress that
the large $\langle{\rm Re}(T)\rangle\sim 32$ required for the quintessence axion
is essential for $M_{GUT}$ to have a value close to the
phenomenologically favored value.
Large ${\rm Re}(T)$ leads also to the length of the 11-th segment 
significantly bigger than the 11-dimensional Planck length:
\beq
\pi\rho\sim 10\kappa^{2/9},
\nonumber 
\eeq
implying that we are indeed in the 11-dimensional heterotic $M$-theory limit 
corresponding to the strong coupling limit of the heterotic
$E_8\times E_8^{\prime}$ string theory.

\medskip

B. {\it  Couplings and Scales in Type I string with D9 and D5-branes}: 

\medskip

Again for simplicity consider
a model  without $T$-dependent threshold correction to $f_9$ in
(\ref{type1gauge}).
We then have \cite{polchinski}
\beqar
&&M_P^2=2(2\pi)^{-7}\alpha^{\prime -4}e^{-2D}V,
\nonumber \\
&&\frac{1}{\alpha_9}=(2\pi)^{-6}\alpha^{\prime -3}e^{-D}V={\rm Re}(S),
\nonumber \\
&&\frac{1}{\alpha_{5i}}=(2\pi)^{-2}\alpha^{\prime -1}e^{-D}
\int_{{\cal C}_i}\omega={\rm Re}(T_i),
\eeqar
and so 
\beqar
&& e^{2D}=\frac{{\rm Re}(S)}{(\sum_{ijk}C_{ijk}/6)[{\rm Re}(T)]^3},
\nonumber \\
&& M_{KK}\equiv \frac{\gamma}{V^{1/6}}
=\frac{\gamma M_P}{
2\sqrt{\pi}(\sum C_{ijk}/6)^{1/6}[{\rm Re}(S)]^{1/2}[{\rm Re}(T)]^{1/2}}
\nonumber \\
&&
M_{string}\equiv\frac{1}{\sqrt{\alpha^{\prime}}}=
\frac{\sqrt{\pi}M_P}{
(\sum C_{ijk}/6)^{1/2}[{\rm Re}(S)]^{1/4}[{\rm Re}(T)]^{3/4}}.
\eeqar
In Type I case with a quintessence axion
${\rm Im}(T)$ or ${\rm Im}(S)$, we also have
\beqar
&&M_{GUT}=M_{KK} \, \, \,  {\rm or} \, \, \, M_{string},
\nonumber \\
&&\frac{1}{\alpha_{GUT}}= {\rm Re}(S) \, \, \,  {\rm or} \, \, \, {\rm Re}(T)
\sim 25,
\nonumber \\
&&\frac{1}{2\pi}\ln(m_{3/2}^2M_P^2/V_Q)={\rm Re}(T) \, \, \, {\rm or} \, \,\,
{\rm Re}(S)\sim 32, 
\nonumber
\eeqar
depending upon how the standard model gauge couplings are embedded \cite{munoz}.
For these moduli VEVs,
\beq
e^{2D}\ll 1, \quad 
\alpha^{\prime}M_{KK}^2={\cal O}(1),
\nonumber
\eeq
indicating that we are in a domain of weakly coupled string but of
strongly coupled worldsheet sigma model.
It is also straightforward to see that
they also give $M_{GUT}$ which is very close to the phenomenologically
favored value, similarly to the case of heterotic $M$-theory.

\medskip

C. {\it Quintessence Axion Scale and the Fine Tuning Problem
of Initial Condition:}

\medskip

Let $K_0(Z+Z^*)$ 
denote the (dimensionless) 
K\"ahler potential of the quintessence axion multiplet
$Z=T$ or $S$.
When all massive moduli (including ${\rm Re}(Z)$) are integrated out,
the  effective lagrangian of the superlight  quintessence axion is given by
\beqar
{\cal L}_{\rm eff}&=&
M_P^2\langle K_0^{\prime\prime}\rangle
[\partial_{\mu} {\rm Im}(Z)]^2
+m_{3/2}^2M_P^2e^{-2\pi \langle {\rm Re}(Z)\rangle}\cos [2\pi {\rm Im}(Z)]
\nonumber \\
&=& \frac{1}{2}(\partial_{\mu} Q)^2+ 
m_{3/2}^2M_P^2e^{-2\pi \langle {\rm Re}(T)\rangle}
\cos (Q/f_Q)
\label{axionlagrangian}
\eeqar
where the canonical 
quintessence axion $Q$ and its decay constant $f_Q$ are given by
\beq
Q=M_P\sqrt{2\langle K_0^{\prime\prime}\rangle}{\rm Im}(Z),
\quad
f_Q=\frac{1}{2\pi}M_P\sqrt{2\langle K_0^{\prime\prime}\rangle}
\nonumber
\eeq
with the K\"ahler metric
$K_0^{\prime\prime}=\partial^2 K_0/\partial Z\partial Z^*$.

In order to have negative pressure,
$Q$ must roll slowly and so satisfy 
\beq
\left|\frac{\partial V_Q}{\partial Q}\right|\ler
\left|\frac{V_Q}{M_P}\right|.
\eeq
When applied for (\ref{axionlagrangian}),
this slow-roll condition leads to 
\bea
&& |2\pi {\rm Im}(Z)|_{\rm present}\ler 
\sqrt{2\langle K_0^{\prime\prime}\rangle}/2\pi, 
\nonumber \\
&& |2\pi{\rm Im}(\dot{Z})|_{\rm present}\ler
H_0/\sqrt{2\langle K_0^{\prime\prime}\rangle},
\label{slowrollcondition}
\eea
implying that at present the angular field $2\pi {\rm Im}(Z)$ 
should be at near
the top of its effective potential 
unless the K\"ahler metric $K_0^{\prime\prime}$ is significantly
bigger than the unity.
(Here $H_0$ denotes the present Hubble expansion rate.)

At leading order approximation in string or $M$ theory,
we have \cite{polchinski}
\beq
K_0\ap -c\ln(Z+Z^*),
\eeq
where $c$ is a constant of order unity.
Obviously this gives 
a small K\"ahler metric 
$K^{\prime\prime}_0\sim 1/[{\rm Re}(Z)]^2\ll 1$ for ${\rm Re}(Z)\sim 32$.
This then leads to
\beq
\left|\frac{\partial^2 V_Q}{\partial Q^2}\right|=
\left|\frac{2\pi^2}{\langle K_0^{\prime\prime}\rangle}\frac{V_Q}{M_P^2}
\right| \gg \left|\frac{V_Q}{M_P^2}\right|,
\eeq
and thus the quintessence axion in string or $M$ theory
largely violates 
the second slow-roll
condition:
\beq
\left|\frac{\partial^2 V_Q}{\partial Q^2}\right|\lesssim 
\left|\frac{V_Q}{M_P^2}\right|.
\label{secondslow1}
\eeq
As a  consequence,
the quintessence axion can lead to an accelerating universe at present,
i.e. can satisfy Eq. (\ref{slowrollcondition}),
{\it only} when it has a very particular initial value in the early universe,
i.e. the quintessence axion suffers from  the  fine-tuning problem
of the initial conditions of $Q$ and $\dot{Q}$
in the early universe. 
As long as
$\sqrt{\langle K_0^{\prime\prime}\rangle}/\pi\ll 1$,
the corresponding quintessence axion requires
the fine tuning of initial condition,
however the degree of required fine tuning is quite sensitive to the value
of $\sqrt{\langle K_0^{\prime\prime}\rangle}$.
Although $K_0^{\prime\prime}\ll 1$ at leading order perturbation theory, 
it can be significantly enhanced by {\it ${\rm Im}(Z)$-independent}
nonperturbative effects which would be responsible
for stabilizing ${\rm Re}(Z)$ at large value $\sim 32$ \cite{banks1}.
In this case,
the fine tuning problem of initial conditions would be ameliorated,
however it is hard to imagine that nonperturbative effects are  strong
enough to give $K_0^{\prime\prime}\sim \pi^2$, so that allow to
avoid the fine tuning of initial condition.

The above discussed fine tuning problem
of initial conditions
appears to be a serious  difficulty
of the quintessence axion.
Here we present a late time inflation scenario
which would resolve this
difficulty for a reasonably enlarged value of
the K\"ahler metric 
$\langle K_0^{\prime\prime}\rangle$.
The gauge symmetries of string or $M$-theory include discrete modular group
\cite{polchinski}
under which $Z$ and 
other generic moduli $\Phi$ transform as
\beq
{\rm Re}(Z)\rightarrow
\frac{1}{{\rm Re}(Z)},
\quad
{\rm Im}(Z)\rightarrow {\rm Im}(Z)+1,
\quad
\Phi\rightarrow\Phi^{\prime},
\eeq
and also CP \cite{gaugecp} under which
\beq
Z\rightarrow Z^*,
\quad
\Phi\rightarrow \Phi^*.
\eeq
Here we take the simplest  form of
the $Z$-duality ($Z=S$ or $T$),
i.e. ${\rm Re}(Z)\rightarrow 1/{\rm Re}(Z)$
with the self-dual value ${\rm Re}(Z)=1$,
however our discussion is valid for
other forms of the $Z$-duality transformation
as long as the self-dual value is of order unity.

The modular and CP invariances  assure that
the invariant points,
\beq
{\rm Re}(Z)=1,
\quad
{\rm Im}(Z)=0 \, \, \, {\rm or} \, \, \, \frac{1}{2},
\quad
\Phi=\Phi^* \, \, \, {\rm or} \, \, \, \Phi^{\prime *},
\eeq
correspond to the stationary points of
the effective action \cite{nir}.
It is then quite possible that
$Z=1$ was  a local minimum
of the effective potential
{\it during the inflation period} 
if the inflaton is  moduli invariant field.
At this inflationary minimum,
all the moduli masses have the same order of magnitude:
\beq
m_{{\rm Re}(Z)}\sim
m_{{\rm Im}(Z)}\sim m_{\Phi}\sim H_{\rm inf},
\eeq
where
$H_{\rm inf}=\sqrt{V_{\rm inf}/M_P^2}$ is the expansion
rate during inflation.
Note that the axion  mass of  $m_{{\rm Im}(Z)}$ 
is comparable to  other moduli masses for 
$\langle {\rm Re}(Z)\rangle={\cal O}(1)$.
The modulus ${\rm Re}(Z)$ can  have  appropriate couplings
to  inflaton field, so that if inflaton  takes its present value,
${\rm Re}(Z)=1$ becomes an unstable point of the moduli
potential. Then after the inflation ${\rm Re}(Z)$ moves 
toward the {\it present minimum}
at ${\rm Re}(Z)\sim 32$
for which
\beq
m_{{\rm Re}(Z)}\sim m_{3/2},
\quad
m_{{\rm Im}(Z)}\sim e^{-\pi\langle {\rm Re}(Z)\rangle} m_{3/2}.
\eeq

About the location of the minimum  in the axion direction, 
we have just two possibilities
if CP is {\it not} spontaneously broken
in the moduli sector.
One of the two CP invariant points,
${\rm Im}(Z)=0$ and $1/2$, is the minimum,
while the other corresponds to the maximum.
Our key assumption on the axion potential is that
${\rm Im}(Z)=0$ is the minimum for
the inflationary modulus value ${\rm Re}(Z)\sim 1$,
however it becomes the maximum for the present value
${\rm Re}(Z)\sim 32$.
Note that  the coefficient of the cosine potential of ${\rm Im}(Z)$
is a function of ${\rm Re}(Z)$, and so its sign
can be changed  when ${\rm Re}(Z)$ varies from 
the inflationary value to the present value.

Given the properties of the moduli potential discussed above,
the cosmological scenario  yielding the present values of
${\rm Im}(Z)$ and
${\rm Im}(\dot{Z})$ satisfying Eq. (\ref{slowrollcondition})
goes as follows.
There was a relatively late  inflationary period  in the early universe
during which ${\rm Re}(Z)$ and other moduli were settled down 
near at the modular and CP-invariant 
local minimum of the  potential, i.e.
at $\langle {\rm Re}(Z)\rangle_{\rm inf}=1$, 
$\langle{\rm Im}(Z)\rangle_{\rm inf}=0$,
and $\langle\Phi\rangle_{\rm inf}=\langle\Phi^*\rangle_{\rm inf}$.
Since the modulus ${\rm Re}(Z)$ is significantly away from the 
present value $\sim 32$, it is expected that the Hubble expansion rate
$H_{\rm inf}$ of this late  inflation is {\it at least} of order $m_{3/2}$.
To avoid a too large quantum fluctuation of
the quintessence axion,  we assume  the minimal value 
\beq
H_{\rm inf}={\cal O}(m_{3/2}).
\eeq
During  the  inflation period, all moduli masses including that of
${\rm Im}(Z)$ are of order $H_{\rm inf}$,
and thus all moduli are settled down at 
the values sufficiently near at the minimum
if the number of inflation efoldings ($N_e$) are large enough \cite{randall}.
In particular, just after the inflation, we have
\bea
&& |2\pi {\rm Im}(Z)|_{\rm inflation}= {\cal O}(e^{-3N_e/2})+{\cal O}( H_{\rm inf}/M_P),
\nonumber \\
&& |2\pi {\rm Im}(\dot{Z})|_{\rm inflation}= {\cal O}(e^{-3N_e/2} H_{\rm inf})+
{\cal O}(H_{\rm inf}^2/M_P),
\label{initialafterinflation}
\eea
where the exponential suppression in the right hand side is 
due to the classical
evolution toward the minimum of $V_{\rm inflation}$, while the second terms
denote the effects of
quantum fluctuations.
After the inflation,
${\rm Re}(Z)$ moves toward the present value ${\rm Re}(Z)\sim 32$.
After ${\rm Re}(Z)$ is settled down at its present value,
${\rm Im}(Z)=0$  becomes the maximum of the axion potential
with
\beq
m_0\equiv \left|\frac{\partial^2 V_Q}{\partial Q^2}\right|^{1/2}_{Q=0}
\approx \frac{\sqrt{3}\pi}{\sqrt{\langle K_0^{\prime\prime}
\rangle}}H_0.
\eeq
Since  $m_0\gg H_0$,
 the inflationary value
${\rm Im}(Z)\sim 0$ is highly unstable  againt the cosmological
evolution  during the period with $H_0\lesssim H\lesssim m_0$.
The resulting  unstability factor is given by $e^{\alpha m_0/H_0}$ where
$\alpha$ is a constant of order unity whose precise value
depends upon the initial
conditions. More detailed study \cite{hclee} suggests $\alpha\approx 0.5$
for the most of
interesting initial conditions.
In order for the quintessence axion to  provide an accelerating 
universe at present,
we then need
\beqar
&&
|2\pi {\rm Im}(Z)|_{\rm inflation}\ler e^{-0.5 m_0/H_0}
\sim e^{-3/\sqrt{\langle K_0^{\prime\prime}\rangle}},
\nonumber \\
&&
|2\pi {\rm Im}(\dot{Z})|_{\rm inflation}\ler H_{\rm inf}
e^{-0.5 m_0/H_0}\sim H_{\rm inf}e^{-3/\sqrt{\langle K_0^{\prime\prime}
\rangle}}.
\eeqar

In order for the initial condition 
(\ref{initialafterinflation}) set by the late time inflation
to satisfy the above condition, 
we need first
a large  efolding:
\beq
N_e \gtrsim  \frac{2}{\sqrt{\langle K_0^{\prime\prime}\rangle}},
\label{largeefolding}
\eeq
and also a small quantum fluctuation of the quintessence axion 
field during (and also after)  the
late inflation:
\beq
\delta Q\sim \frac{H_{\rm inf}}{2\pi}\sim
\frac{m_{3/2}}{2\pi}\ler M_P e^{-3/\sqrt{\langle K_0^{\prime\prime}\rangle}}.
\label{quantum}
\eeq
In fact, $N_e$ can not be arbitrarily large.
For late time inflation with $H_{\rm inf}\sim m_{3/2}$, it is very difficult
to generate  the observed density fluctuation $\delta\rho/\rho
\sim 10^{-5}$. It  is thus reasonable to assume that density fluctuations
were created before the late inflation. Then the late inflation
is required not to destroy the pre-exisiting density fluctuations.
This consideration gives an upper bound of $N_e$ \cite{randall},
yielding
\beq
N_e\ler 25\sim 30
\label{smallefolding}
\eeq
for a weak scale $m_{3/2}$ and the reheat temperature range
$T_r=10^5\sim 10^{-2}$ GeV. 

The above conditions (\ref{largeefolding}),
(\ref{quantum}) and (\ref{smallefolding})
can not be simultaneously satisfied 
for the perturbative value of $\langle K_0^{\prime\prime}\rangle$.
At leading order in perturbation theory, we have
$K_0\sim 3 \ln (Z+Z^*)$ and thus 
$\sqrt{\langle K_0^{\prime\prime}\rangle}\sim 1/\langle {\rm Re}(Z)\rangle\sim
3\times 10^{-2}$ for $\langle {\rm Re}(Z)\rangle\sim 32$,
which is obvioulsy in conflict with the above conditions.
However as was noted in the  \cite{banks1}, 
there can be a rather large {\it nonperturbative}
contribution to the K\"ahler potential 
which may be essential for  stabilizing ${\rm Re}(Z)$ at 
the desired value $\sim 32$.
In particular, if the nonperturbative terms involve a large power of
$(Z+Z^*)$, e.g. $K_{np}=h(Z+Z^*)^ke^{-b\sqrt{Z+Z^*}}={\cal O}(1)$ 
with $k\gtrsim 6$, the K\"ahler metric can be significantly
enlarged to a value
\beq
\sqrt{\langle K_0^{\prime\prime}\rangle}\approx 9\times 10^{-2}
\eeq
which would  accomodate all the conditions for the quintessence
axion to lead to an accelerating universe at present.

\medskip

{\bf Acknowledgments}:
I thank K. Hwang for drawing the figures and H. B. Kim and S. Thomas for
useful discussions.
This work is supported in part
by KOSEF Grant 981-0201-004-2,
KOSEF through CTP of Seoul National University,
KRF under the Grant 1998-015-D00071 and the Distinguished Scholar Exchange Program.

\font\thinlinefont=cmr5
\begingroup\makeatletter\ifx\SetFigFont\undefined
\def\x#1#2#3#4#5#6#7\relax{\def\x{#1#2#3#4#5#6}}%
\expandafter\x\fmtname xxxxxx\relax \def\y{splain}%
\ifx\x\y   
\gdef\SetFigFont#1#2#3{%
  \ifnum #1<17\tiny\else \ifnum #1<20\small\else
  \ifnum #1<24\normalsize\else \ifnum #1<29\large\else
  \ifnum #1<34\Large\else \ifnum #1<41\LARGE\else
     \huge\fi\fi\fi\fi\fi\fi
  \csname #3\endcsname}%
\else
\gdef\SetFigFont#1#2#3{\begingroup
  \count@#1\relax \ifnum 25<\count@\count@25\fi
  \def\x{\endgroup\@setsize\SetFigFont{#2pt}}%
  \expandafter\x
    \csname \romannumeral\the\count@ pt\expandafter\endcsname
    \csname @\romannumeral\the\count@ pt\endcsname
  \csname #3\endcsname}%
\fi
\fi\endgroup

\begin{figure}
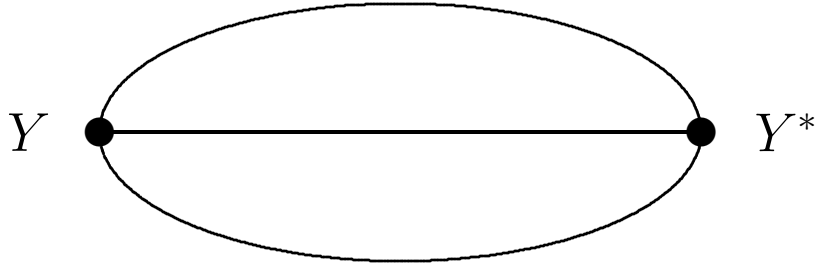

{
\centering
\vspace*{15mm}
\mbox{\beginpicture
\setcoordinatesystem units <0.90000cm,0.90000cm>
\unitlength=0.90000cm
\linethickness=1pt
\setplotsymbol ({\makebox(0,0)[l]{\tencirc\symbol{'160}}})
\setshadesymbol ({.})
\setlinear
%
%
\linethickness=1pt
\setplotsymbol ({\makebox(0,0)[l]{\tencirc\symbol{'160}}})
\ellipticalarc axes ratio  4.445:1.905  360 degrees 
	from 10.795 22.225 center at  6.350 22.225
%
%
\linethickness=1pt
\setplotsymbol ({\makebox(0,0)[l]{\tencirc\symbol{'160}}})
\put{\makebox(0,0)[l]{\circle*{ 0.449}}} at  1.905 22.225
%
%
\linethickness=1pt
\setplotsymbol ({\makebox(0,0)[l]{\tencirc\symbol{'160}}})
\put{\makebox(0,0)[l]{\circle*{ 0.449}}} at 10.795 22.225
%
%
\linethickness=1pt
\setplotsymbol ({\makebox(0,0)[l]{\tencirc\symbol{'160}}})
\putrule from  1.905 22.225 to 10.795 22.225
%
%
\put{\SetFigFont{20}{34.8}{rm}$Y$} [c] at  0.859 22.225
%
%
\put{\SetFigFont{20}{34.8}{rm}$Y^*$} [c] at 12.048 22.225
\linethickness=0pt
\putrectangle corners at  0.159 24.160 and 11.748 20.290
\endpicture}

\medskip

\medskip

\caption{
Two loop supergraph inducing the quadratically-divergent potential energy of
$Y_{\alpha\beta\gamma}$. 
}
} \end{figure}

\bigskip

\begin{figure}
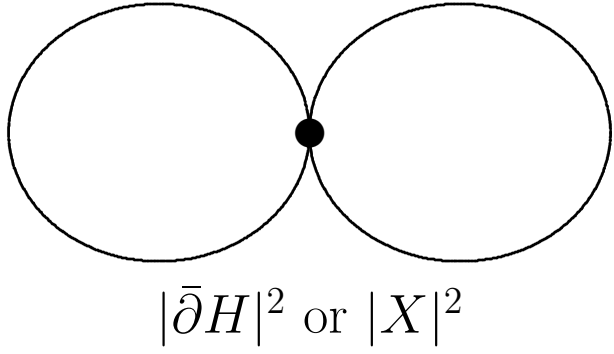

{
\centering
\mbox{\beginpicture
\setcoordinatesystem units <0.90000cm,0.90000cm>
\unitlength=0.90000cm
\linethickness=1pt
\setplotsymbol ({\makebox(0,0)[l]{\tencirc\symbol{'160}}})
\setshadesymbol ({\thinlinefont .})
\setlinear
%
%
\linethickness=1pt
\setplotsymbol ({\makebox(0,0)[l]{\tencirc\symbol{'160}}})
\ellipticalarc axes ratio  2.223:1.905  360 degrees 
	from 10.795 22.225 center at  8.572 22.225
%
%
\linethickness=1pt
\setplotsymbol ({\makebox(0,0)[l]{\tencirc\symbol{'160}}})
\ellipticalarc axes ratio  2.223:1.905  360 degrees 
	from  6.350 22.225 center at  4.128 22.225
%
%
\linethickness=1pt
\setplotsymbol ({\makebox(0,0)[l]{\tencirc\symbol{'160}}})
\put{\makebox(0,0)[l]{\circle*{ 0.449}}} at  6.350 22.225
%
%
\put{\SetFigFont{20}{34.8}{rm}$|\bar{\partial}H|^2$ or $|X|^2$} [c] at 6.350
19.5
\linethickness=0pt
\putrectangle corners at  1.873 24.160 and 10.827 18.891
\endpicture}

\medskip

\medskip

\caption{
Quartically-divergent two loop supergraph for the potential energy of
$H_{\alpha\beta}$ and $X_{\alpha\beta\bar\gamma}$.
}
} \end{figure}

\bigskip

\begin{figure}
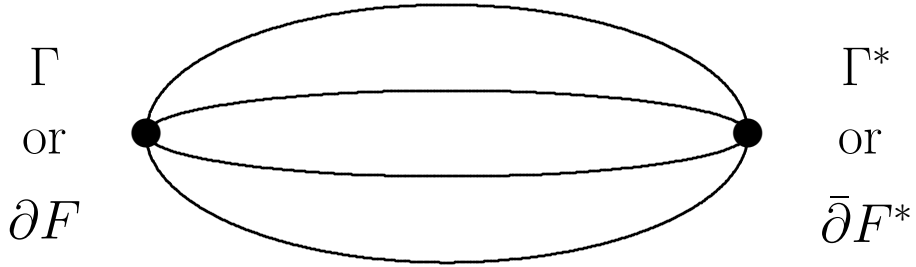

{
\centering
\mbox{\beginpicture
\setcoordinatesystem units <0.90000cm,0.90000cm>
\unitlength=0.90000cm
\linethickness=1pt
\setplotsymbol ({\makebox(0,0)[l]{\tencirc\symbol{'160}}})
\setshadesymbol ({\thinlinefont .})
\setlinear
%
%
\linethickness=1pt
\setplotsymbol ({\makebox(0,0)[l]{\tencirc\symbol{'160}}})
\put{\makebox(0,0)[l]{\circle*{ 0.449}}} at  1.905 22.225
%
%
\linethickness=1pt
\setplotsymbol ({\makebox(0,0)[l]{\tencirc\symbol{'160}}})
\put{\makebox(0,0)[l]{\circle*{ 0.449}}} at 10.795 22.225
%
%
\linethickness=1pt
\setplotsymbol ({\makebox(0,0)[l]{\tencirc\symbol{'160}}})
\ellipticalarc axes ratio  4.445:1.905  360 degrees 
	from 10.795 22.225 center at  6.350 22.225
%
%
\linethickness=1pt
\setplotsymbol ({\makebox(0,0)[l]{\tencirc\symbol{'160}}})
\ellipticalarc axes ratio  4.445:0.635  360 degrees 
	from 10.795 22.225 center at  6.350 22.225
%
%
\put{\SetFigFont{20}{34.8}{rm}$\Gamma$} [c] at  0.4 23.219
%
%
\put{\SetFigFont{20}{34.8}{rm}or} [c] at  0.4 22.107
%
%
\put{\SetFigFont{20}{34.8}{rm}$\partial F$} [c] at  0.4 20.956
%
%
\put{\SetFigFont{20}{34.8}{rm}$\Gamma^*$} [c] at 12.548 23.219
%
%
\put{\SetFigFont{20}{34.8}{rm}or} [c] at 12.448 22.107
%
%
\put{\SetFigFont{20}{34.8}{rm}$\bar\partial F^*$} [c] at 12.548 20.956
\linethickness=0pt
\putrectangle corners at  0.159 24.160 and 11.748 20.290
\endpicture}

\vspace*{13mm}

\caption{
Three loop supergraph for the potential energy of
$\Gamma_{\alpha\beta\gamma\delta}$ and $F_{a\alpha\beta}.$
For $F_{a\alpha\beta}$, two lines denote the gauge multiplets
while the others are the charged  matter multiplets.}
}
\end{figure}


\begin{thebibliography}{99}

\bibitem{accel} S. Perlmutter {\it et al.,} Nature {\bf 391}, 51 (1998);
A. G. Riess {\it et al.,} astro-ph/9805201.


\bibitem{weinberg} For a review, see S. Weinberg, Rev. Mod. Phys.
{\bf 61}, 1 (1989).
 
\bibitem{coleman} S. Coleman, Nucl. Phys. {\bf B310}, 643 (1988). 

\bibitem{quint} M. \"{O}zer and M. O. Taha, Nucl. Phys. {\bf B287},
797 (1987); B. Ratra and P. J. E. Peebles, Phys. Rev. {\bf D37},
3406 (1988); J. A. Frieman, C. T. Hill and R. Watkins, Phys. Rev.
{\bf D46}, 1226 (1992); M. S. Turner and M. White, Phys. Rev.
{\bf D56}, 4439 (1997);
T. Chiba, N. Sugiyama and T. Nakamura, Mon. Not. R. Astron.
Soc. {\bf 289}, 5 (1997); {\it ibid.} {\bf 301}, 72 (1998);
R. R. Caldwell, R. Dave and P. J. Steinhardt,
Phys. Rev. Lett. {\bf 80}, 1582 (1998);
L. Wang and P. J. Steinhardt, astro-ph/9804015;
G. Huey {\it et al.}, astro-ph/9804285;
I. Zlater, L. Wang and P. J. Steinhardt,
astro-ph/9807002; D. Huterer and M. S. Turner,
astro-ph/9808133, T. Chiba and T. Nakamura, Prog. Theor. Phys.
{\bf 100}, 1077 (1998); T. Nakamura and T. Chiba, astro-ph/9810447.

\bibitem{lyth} C. Kolda and D. H. Lyth, hep-ph/9811375.

\bibitem{horava} P. Horava and E. Witten, Nucl. Phys. {\bf B460},
506 (1996); Phys. Rev. {\bf D54}, 7561 (1996).

\bibitem{polchinski} J. Polchinski, {\it String Theory}  (Cambridge University Press, 1998).

\bibitem{banks1} T. Banks and M. Dine, Phys. Rev. {\bf D50}, 7454 (1994);
J. A. Casas, Phys. Lett. {\bf B384}, 103 (1996); K. Choi, H. B. Kim
and H. D. Kim, Mod. Phys. Lett. {\bf A14}, 125 (1999).

\bibitem{nillesreview} For a review, see H. P. Nilles, Phys. Rep. {\bf 110}, 1 (1984).


\bibitem{gaugemediation} For a review, see G. F. Giudice and R. Rattazzi,
hep-ph/9801271.

\bibitem{sugraoneloop} M. T. Grisaru, M. Rocek and A. Karlhede, Phys. Lett.
{\bf B120}, 110 (1983).

\bibitem{bagger} K. Choi, J. E. Kim, and H. P. Nilles, Phys. Rev. Lett. {\bf 73},
1758 (1994); J. Bagger, E. Poppitz and L. Randall, Nucl. Phys. {\bf B455}, 59 (1995).

\bibitem{choileemunoz} M. K. Gaillard, V. Jain and K. Saririan,
Phys. Lett. {\bf B387}, 520 (1996);
K. Choi, J. S. Lee, and C. Munoz, Phys. Rev. Lett. {\bf 80},
3686 (1998).


\bibitem{kimreview} J. E. Kim, Phys. Rep. {\bf 150}, 1 (1987);
H. Y. Cheng, Phys. Rep. {\bf 158}, 1 (1988);
R. D. Peccei, in {\it CP Violation}, ed. C. Jarlskog (World Scientific, 1989).

\bibitem{hyungdo} See for instance K. Choi and H. D. Kim, hep-ph/9809286.

\bibitem{carroll} S. M. Carroll, astro-ph/9806099.


\bibitem{kim} J. E. Kim, hep-ph/9811509.

\bibitem{upquark} H. Georgi and I. N. McArthur, HUTP 81/A011 (1981)
(unpublished); D. B. Kaplan and A. V. Manohar, Phys. Rev. Lett. {\bf 56},
2004 (1986); K. Choi, C. W. Kim and W. K. Sze, Phys. Rev. Lett. {\bf 61},
794 (1988).

\bibitem{binetruy} P. Binetruy, hep-ph/9810553.
 
\bibitem{holman} R. Holman {\it et al.}, Phys. Lett. {\bf B282}, 132 (1992);
M. Kamionkowski and J. March Russel, {\it ibid.} {\bf 282}, 137 (1992);
S. M. Barr and D. Seckel, Phys. Rev.  {\bf D46}, 539 (1992).
 
\bibitem{nilles} H. P. Nilles, Phys. Lett. {\bf B180}, 240 (1986).
 
\bibitem{banks} T. Banks and M. Dine, Nucl. Phys. {\bf B479}, 173 (1996).

\bibitem{nilles1} H. P. Nilles and S. Stieberger, Nucl. Phys. {\bf B499},
3 (1997).

\bibitem{maxion} K. Choi, Phys. Rev. {\bf D56}, 6588 (1997).

\bibitem{benakli} K. Benakli, hep-th/9805181.

\bibitem{stieberger} S. Stieberger, hep-th/9807124.

\bibitem{choikim} K. Choi and J. E. Kim, Phys. Lett. {\bf B165}, 71 (1985);
L. E. Ibanez and H. P. Nilles, Phys. Lett. {\bf B169}, 354 (1986);
J. P. Derendinger, L. E. Ibanez and H. P. Nilles, Nucl. Phys. {\bf B267},
365 (1986).


\bibitem{witten} E. Witten, Nucl. Phys. {\bf B471}, 135 (1996).

\bibitem{cremmer} E. Cremmer {\it et al.}, Nucl. Phys. {\bf B250}, 385 (1985).

\bibitem{curio} J. Louis and K. F\"orger, hep-th/9611184;
G. Curio, hep-th/9708009;
D. L\"ust, hep-th/9803072.

\bibitem{hdkim} K. Choi, H. B. Kim and H. D. Kim, hep-th/9808122,
Mod. Phys. Lett. {\bf A14}, 125 (1999).


\bibitem{worldsheetinstanton} M. Dine, N. Seiberg, X.-G. Wen, and E. Witten,
Nucl. Phys. {\bf B289}, 319 (1987); {\bf B278}, 769 (1986).

\bibitem{worldsheetinstanton1} L. J. Dixon, V. S. Kaplunovsky and J. Louis,
Nucl. Phys. {\bf B355}, 649 (1991); S. Hosono, A. Klemm, S. Theisen
and S.-T. Yau, Nucl. Phys. {\bf B433}, 501 (1995).

\bibitem{munoz} For a recent discussion of Type I string phenomenology
with D-brane configurations,
see L. E. Ibanez, C. Munoz and S. Rigolin, hep-ph/9812397.



\bibitem{gaugecp} K. Choi, D. B. Kaplan and A. E. Nelson,
Nucl. Phys. {\bf B391}, 515 (1993); M. Dine, R. G. Leigh and
D. A. MacIntire, Phys. Rev. Lett. {\bf 69}, 2030 (1992).

\bibitem{nir} M. Dine, Y. Nir and Y. Shadmi, hep-th/9806124.





\bibitem{randall} L. Randall and S. Thomas, Nucl. Phys. {\bf B449}, 229 (1995).

\bibitem{hclee} K. Choi, H. D. Kim and H. C. Lee, in preparation.


\end{thebibliography}
\end{document}